\documentclass[10pt,a4paper]{article}
\usepackage[T1]{fontenc}       
\usepackage{amssymb}
\usepackage{amsmath}
\usepackage{braket}
\usepackage{graphicx}
\usepackage{multicol}
\usepackage{scrextend}
\usepackage{lineno}

\usepackage[sort&compress]{natbib}
\setcitestyle{numbers}
\begin{document}
\begin{center}

{\large{\bf
Spin filling and orbital structure of the first six holes in a silicon metal-oxide-semiconductor quantum dot
}}


\vskip0.5\baselineskip

{\bf
S. D. Liles$^{1}$, R. Li$^{1,3}$, C. H. Yang$^{2}$, F. E. Hudson$^{2}$, M. Veldhorst$^{2,3}$, A. S. Dzurak$^{2}$, A. R. Hamilton$^{1}$
}

\vskip0.5\baselineskip
{\em
$^{1}$School of Physics, University of New South Wales, Sydney NSW 2052, Australia\\
$^{2}$Centre for Quantum Computation and Communication Technology, School of Electrical Engineering and Telecommunications, The University of New South Wales, Sydney NSW 2052, Australia\\ 
$^{3}$QuTech and Kavli Institute of Nanoscience, TU Delft, 2600 GA Delft, The Netherlands\\
}
\end{center}

\begin{multicols}{2}
\textbf{The spin states of electrons confined in semiconductor quantum dots form a promising platform for quantum computation \cite{loss1998quantum,hanson2007spins,zwanenburg2013silicon}. Recent studies of silicon CMOS qubits have shown coherent manipulation of electron spin states with extremely high fidelity \cite{veldhorst2014addressable}. However, manipulation of single electron spins requires large oscillatory magnetic fields to be generated on-chip, making it difficult to address individual qubits when scaling up to multi-qubit devices \cite{koppens2006driven,veldhorst2014addressable}. The spin-orbit interaction allows spin states to be controlled with electric fields, which act locally and are easier to generate\cite{golovach2006electric,flindt2006spin,nowack2007coherent}. While the spin-orbit interaction is weak for electrons in silicon, valence band holes have an inherently strong spin-orbit interaction. However, creating silicon quantum dots in which a single hole can be localised, in an architecture that is suitable for scale-up to a large number of qubits, is a challenge\cite{li2015pauli,spruijtenburg2013single,yamaoka2017charge,betz2014ambipolar}. Here we report a silicon quantum dot, with an integrated charge sensor, that can be operated down to the last hole. We map the spin states and orbital structure of the first six holes, and show they follow the Fock-Darwin spectrum \cite{fock1928bemerkung,darwin1931diamagnetism}. We also find that hole-hole interactions are extremely strong, reducing the two-hole singlet-triplet splitting by $90\%$ compared to the single particle level spacing of 3.5 meV. These results provide a route to single hole spin quantum bits in a planar silicon CMOS architecture.}
	
\begin{figure*}[t]
	\centering
	\includegraphics[width=\textwidth]{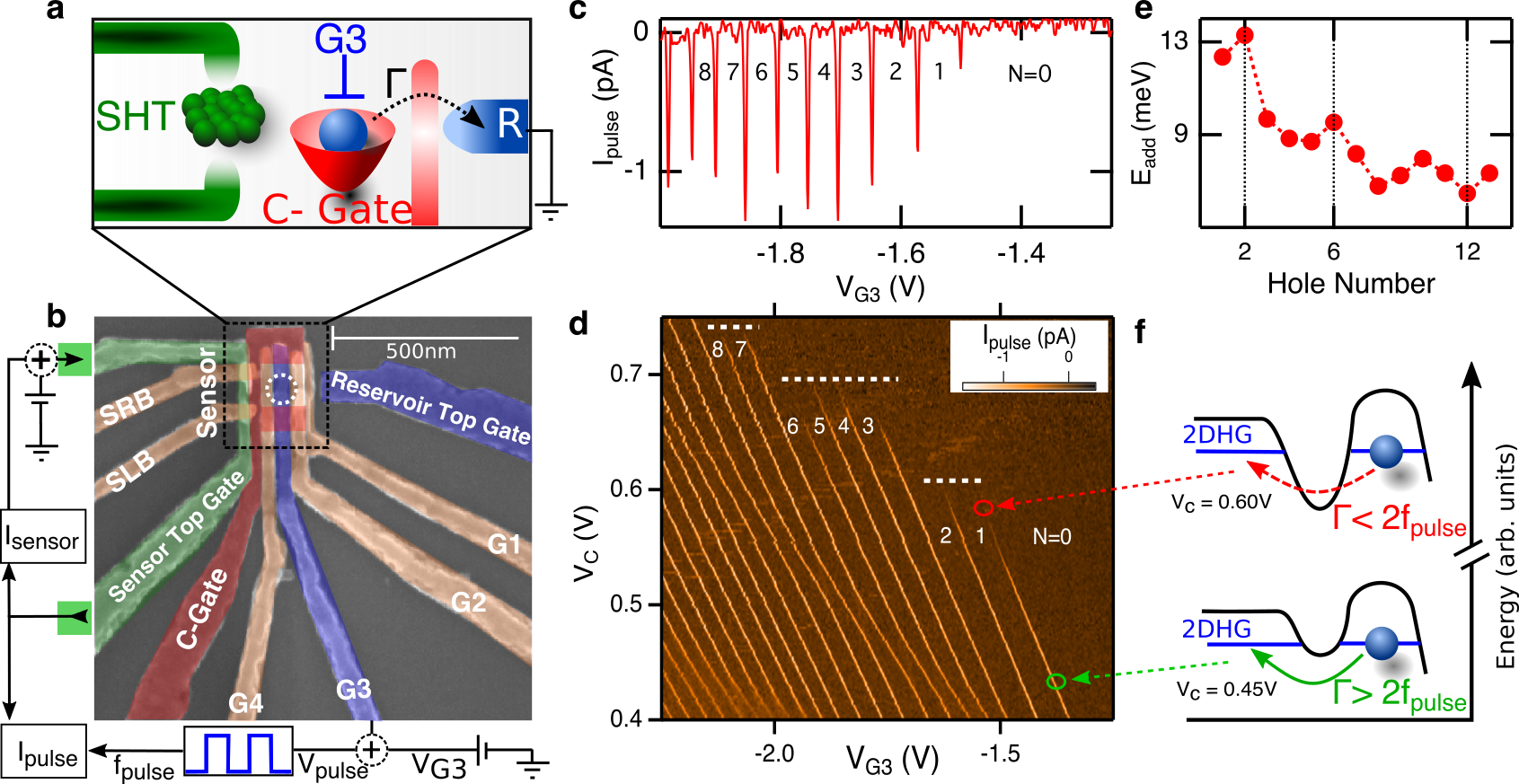} 
	\caption{\textbf{Silicon quantum dot with charge sensor, capable of reaching the last hole.} \textbf{a}, Schematic of the device concept. The device consists of a quantum dot coupled to a single reservoir (R), with an adjacent single hole transistor charge sensor (SHT). The tunnel rate between the dot and reservoir ($\Gamma$) can be tuned using the C-gate voltage ($V_{C}$), and the dot occupation can be controlled with gate G3 voltage ($V_{G3}$). \textbf{b}, False-coloured scanning electron microscope image of an identical device, with the measurement schematic. \textbf{c}, Depletion of the last 10 holes in the quantum dot, showing the $V_{pulse}$ induced signal on the charge sensor measured at $V_{C}$ = 0.47V. \textbf{d}, Charge stability diagram, showing the number of holes on the dot as a function of the confining gate and pulse gate potentials.  The horizontal white lines highlight the disappearance of the charge transition signals as $V_C$ is increased. The transitions signals disappear in distinct groupings, indicating shell filling. Measurements performed for $V_{pulse}=3$mV and $f_{pulse}=333$Hz. A slight bending in the lines in the vicinity of $(V_{G3}=-2$V,$V_C=0.55)$V is due to coupling to nearby confined charge.\textbf{e}, The hole addition energy extracted from (\textbf{d}), showing peaks at $N$=2 and 6 consistent with shell filling. \textbf{f}, Schematic diagram showing how the change in the tunnel barrier with $V_C$ causes the charge transition signals to disappear with increasing $V_C$, when the tunnel time becomes comparable to the length of the pulses applied to gate G3.}
	\label{fig:FigureDevice}
\end{figure*}
In this work, we present observations of the first six hole states in a surface-gated silicon metal-oxide-semiconductor quantum dot. Previous studies of planar silicon-based hole quantum dots have used transport measurements to study the addition spectrum of the quantum dot\cite{li2015pauli,spruijtenburg2013single,yamaoka2017charge}. However, as these devices approach the few hole regime, the tunnel barriers become extremely opaque, and the transport signal falls precipitously. This has hampered studies of hole quantum dots containing one and two holes, which is the most widely used regime for spin-based quantum computation applications. Figure \ref{fig:FigureDevice}a shows a schematic of the device under study in this letter, and Figure \ref{fig:FigureDevice}b shows a scanning electron microscope image of an identical device to the one used in this study. The layout of this device is suitable for high frequency spin manipulation experiments\cite{nowack2007coherent}, and is optimized for scalability up to many qubits \cite{dennis2002topological,jones2016logical}. This device is a hole quantum dot connected to a single reservoir (R) of two-dimensional holes, with an adjacent charge sensor (SHT). By measuring the charge occupation with a charge sensor we are able to study hole states even when the tunnel rate between the dot and reservoir is much smaller than could be detected in transport. In this experiment the number of holes on the dot $N$ is controlled with the bias on gate G3. The dot-reservoir tunnel rate $\Gamma$ can be tuned without affecting the dot confinement shape using the bias applied to the C-gate. 

Hole spins in semiconductor quantum dots are attracting significant attention as candidates for spin-based quantum computation applications \cite{szumniak2012spin,kloeffel2013circuit}, with recent experiments exploiting the strong spin-orbit coupling to demonstrate all-electrical spin manipulation using gate electrodes \cite{pribiag2013electrical,maurand2016cmos}. Hole spins have the benefit that they are rather insensitive to dephasing induced by hyperfine coupling to the host crystals nuclei-spin\cite{bulaev2005spin,keane2011resistively}. This source of dephasing is the leading cause of decoherence in electron spin qubits\cite{koppens2008spin}. Although the effects of hyperfine induced dephasing can be minimized using experimental\cite{viola1998dynamical} or material\cite{veldhorst2014addressable} based techniques, the weak coupling of holes to nuclear spins provides the possibility for long coherence times without the need for additional experimental complexity\cite{hu2012hole,higginbotham2014hole}. Further, there is no valley degeneracy in the silicon valence band, avoiding complexities that affect electrons in silicon quantum dots\cite{yang2012orbital}. Despite these promising properties, hole based quantum dots still face technological challenges that have been overcome in electron systems more than a decade ago\cite{zwanenburg2013silicon}. Demonstrating the ability to isolate one hole and determining the spin properties of the last few holes in surface gated silicon MOS based quantum dots would be a key advance towards realizing hole-spin-based qubits.

In order to characterize the addition spectrum of holes in the quantum dot we employ a pulse-bias technique\cite{elzerman2004excited}, which allows the the charge occupation of the dot to be monitored using an adjacent charge sensor (SHT). We apply a 1mV DC excitation to the SHT's source ohmic contact, and add a continuous square wave of magnitude $V_{pulse}$ and frequency $f_{pulse}$ to gate G3.  The modulation of the DC sensor current by $V_{pulse}$, called $I_{pulse}$, is sensitive to $dQ_{dot}/dV_{G3}$ (as long as $\Gamma > 2f_{pulse}$). In Figure \ref{fig:FigureDevice}c we show a measurement of $I_{pulse}$ as $V_{G3}$ is swept. At specific values of $V_{G3}$ a hole is able to tunnel on and off the dot during the positive/negative phase of $V_{pulse}$. This charge movement decreases the DC sensor current, causing a  negative spike in $I_{pulse}$ of width $V_{pulse}$ in the $V_{G3}$ scan. The measurement of Figure \ref{fig:FigureDevice}c was repeated over a range of $V_{C}$ to produce the charge stability diagram in Figure \ref{fig:FigureDevice}d. The identification of the last hole in the dot is confirmed by the absence of any additional charge transitions beyond the region labeled N=0 in Figure \ref{fig:FigureDevice}d.

The spacing of the charge transition lines in Figure \ref{fig:FigureDevice}d provides clear evidence for orbital shell filling of the hole quantum dot. We extracted the addition energy ($E_{add}(N) =  \mu_{N+1} - \mu_{N}$) by measuring the spacing $\Delta V_{G3}$ between consecutive Coulomb peaks, then converted $\Delta V_{G3}$ to energy using the lever arm of 0.17 eV/V (see Supporting Information). In Figure \ref{fig:FigureDevice}e we plot the addition energy $E_{add}$ for increasing hole number. A clear increase in the addition energy is observed for N=2 and N=6, which suggests the orbital shell is full for the second and sixth holes.

Further evidence for orbital shell filling is given by the stair-like disappearance of charge transition signals, which is highlighted by the dashed horizontal white lines in Figure \ref{fig:FigureDevice}d. Along each vertical charge transition line the measured signal decreases as $V_{C}$ is made more positive. As $V_{C}$ becomes more  positive the tunnel barrier becomes more opaque, and subsequently the tunnel rate from the dot to the reservoir $\Gamma$ decreases. The charge sensor transition is no longer visible  when $\Gamma < 2f_{pulse}$, as shown schematically in Figure \ref{fig:FigureDevice}f. When a hole in the dot occupies a higher energy orbital shell, its wavefunction span increases, which increases the tunnel rate. Hence, the charge sensor transition signals should lose visibility at higher $V_{C}$ for holes in higher orbitals. We observe that the N=(1,2), (3,4,5,6) and (7,8) charge transition lines disappear at the same $V_{C}$ (dashed lines in Figure \ref{fig:FigureDevice}d), suggesting that these holes fill the same orbital state, with similar tunnel rates in the same orbital level.

These observations of shifts in addition energy and tunnel rate suggest the first two holes fill into the first orbital, and the next four holes fill into the second orbital. This shell filling is consistent with the Fock-Darwin orbital structure for a 2D parabolically confined quantum dot\cite{fock1928bemerkung,darwin1931diamagnetism}. Beyond N=6 the observed orbital filling departs from the so-called 2D magic numbers, which may reflect a loss of circular symmetry of the parabolic confinement for higher hole occupation, or many-body effects\cite{ciorga2000addition}.

\begin{figure*}[t]
	\centering
	\includegraphics[scale=0.4]{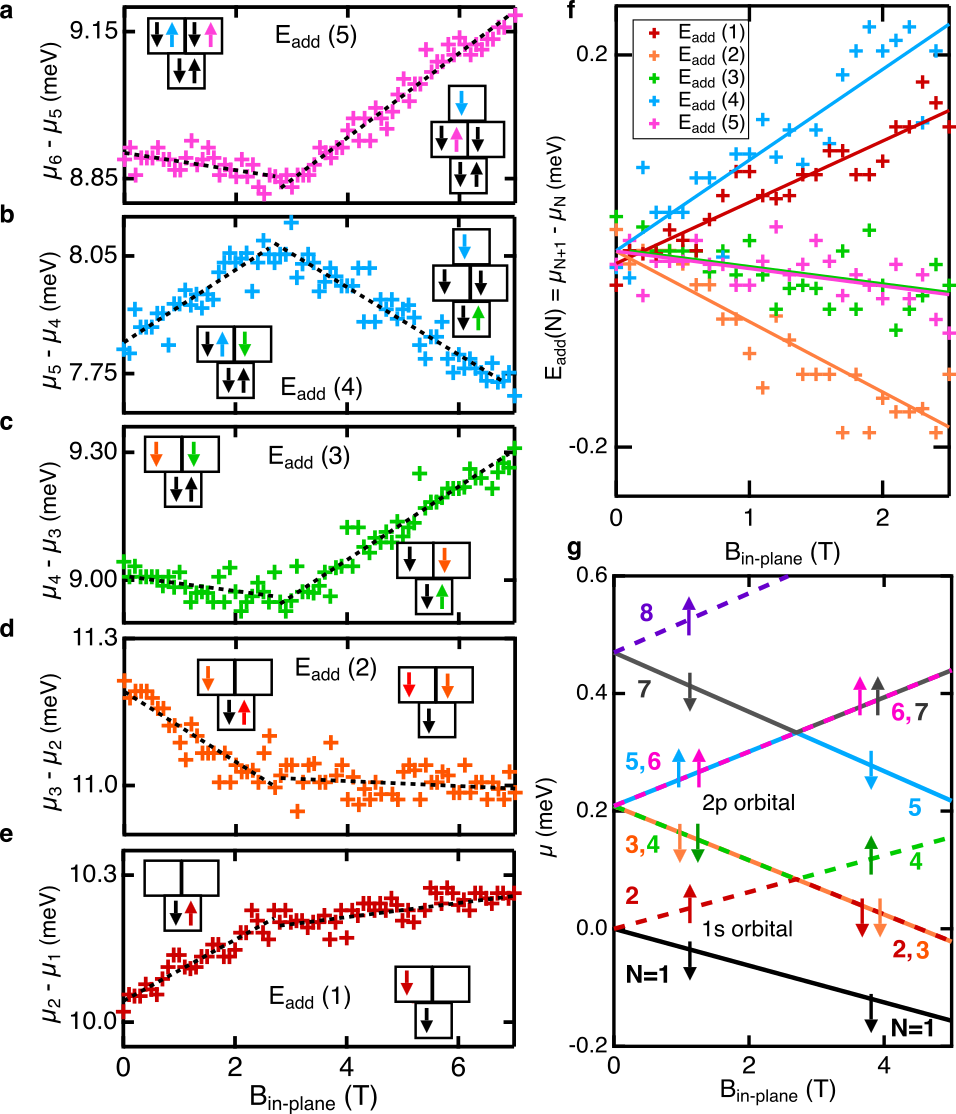}
	\caption{\textbf{Spin filling sequence and orbital structure.} \textbf{a}-\textbf{e}, Addition energy for the first 6 holes as a function of in-plane magnetic field. The black dashed line is a linear fit to the raw data over the region |B|<2.7T (low field) and |B|>2.7T (high field). The left and right inset shows the infered ground state spin filling for the low and high magnetic field regions respectively. \textbf{f}, The addition energies of (\textbf{a})-(\textbf{e}) plotted over the low field region with data offset to clarify the three distinct slopes, positive, negative, and close to zero. Solid lines are least squares fit to the data. \textbf{g}, Model of the hole orbital shell structure for the first eight holes (ignoring Coulomb charging energy), which explains the observed addition energies in (\textbf{a})-(\textbf{e}). The colours of the orbitals corespond to the hole charge occupations in (\textbf{a})-(\textbf{e}).}
	\label{fig:ShellFilling}
\end{figure*}
To understand the spin structure of the hole quantum dot, we study the magnetic field dependence of the addition energy of the hole dot for N=1 to 6 holes. In Figures \ref{fig:ShellFilling}a$-$e we show the addition energy $\mu_{n+1}-\mu_{n}$ for the first six holes as a function of in-plane magnetic field B. The slope of the N$^{th}$ addition energy $dE_{add}/dB$ with respect to B depends on the relative spin orientation of the (N+1)$^{th}$ and N$^{th}$ hole, with three distinct possibilities:
\begin{equation}
\begin{array}{lcl}
&+g^{*}\mu_{B} &\downarrow \uparrow \\
\frac{dE_{add}}{dB} = &-g^{*}\mu_{B} &\uparrow \downarrow \\
&0 &  \uparrow \uparrow or \downarrow \downarrow \\
\end{array}
\end{equation}
where the first and second arrow depicts the spin filling sequence of the (N+1)$^{th}$ and N$^{th}$ holes respectively. We refer to |B|<2.7T as the low field region, and |B|>2.7T as the high field region. In both the low and high field region of Figures \ref{fig:ShellFilling}a$-$e the slope $dE_{add}/dB$ is either positive, negative or close to zero, as shown by the dashed lines. Figure \ref{fig:ShellFilling}f shows that the slope of the addition energy for the first six holes in the low field region takes one of three distinct values, consistent with equation 1.

In Figures \ref{fig:ShellFilling}a$-$e we observe a change in slope at 2.7T. This change in slope suggests a change in the spin filling sequence, which is a result of magnetic field induced orbital level crossings. Knowing the  hole number and the relative spin orientations, we can build up the spin shell filling structure in both the low and high field regimes, as indicated in the left and right insets of Figures \ref{fig:ShellFilling}a$-$e (see Supporting Information).

We now discuss the spin filling sequence in detail, beginning with the low field spin filling. The first and second holes form a Pauli spin pair in a two-fold degenerate orbital, labeled 1\textit{s}.  The third and fourth holes fill the 2\textit{p$_{x}$} and 2\textit{p$_{y}$} states with spins parallel to each other. The fifth and sixth holes fill the 2\textit{p$_{x}$} and 2\textit{p$_{y}$} states with spins parallel to each-other, but opposite to the third and fourth holes.

\begin{figure*}[t]
	\centering
	\includegraphics[scale=0.3]{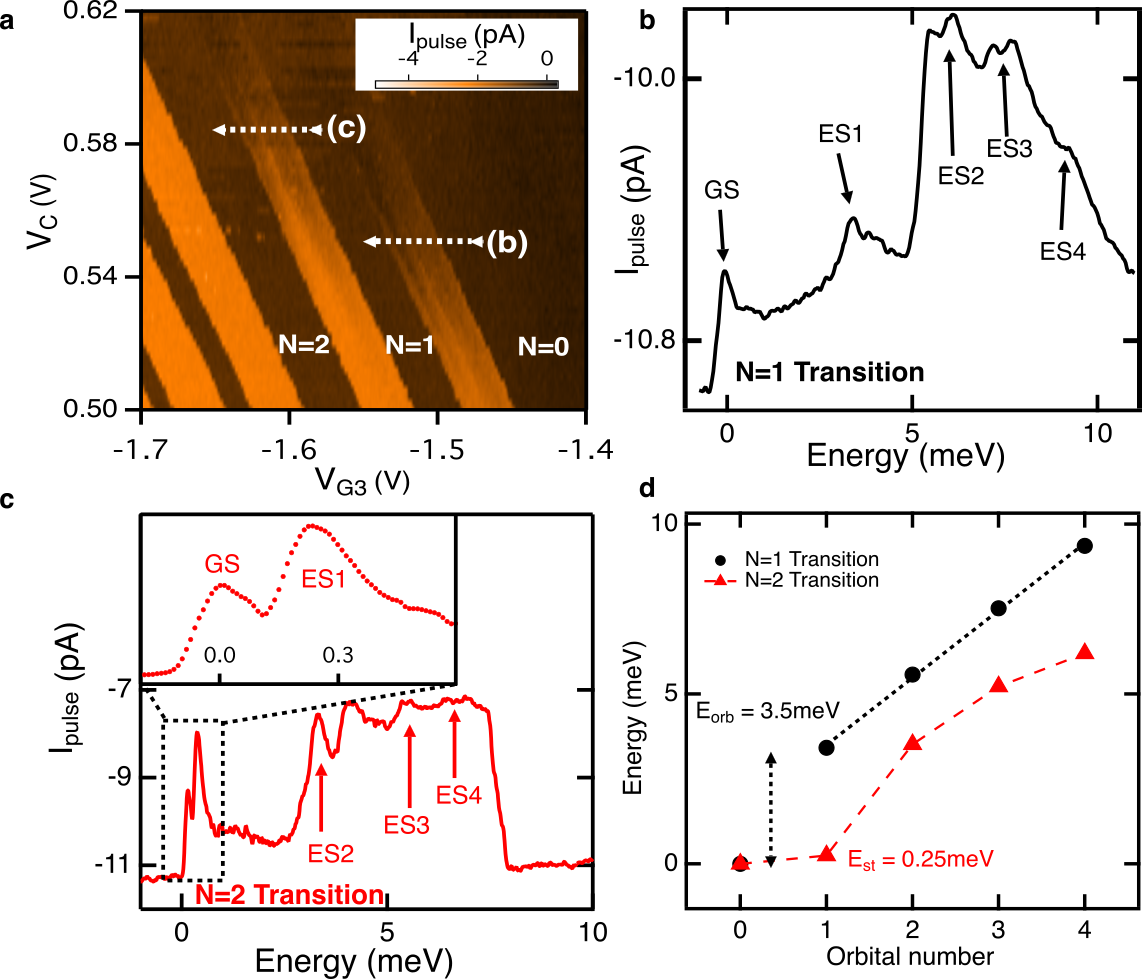}
	\caption{\textbf{Excited state spectroscopy.} \textbf{a}, Charge stability diagram for $V_{pulse} = 40$mV and $f_{pulse}=333$Hz. The white dashed lines labelled (b) and (c) corespond to cut taken to obtain the data in \textbf{b} and \textbf{c} respectively. \textbf{b}, Measurement of $I_{pulse}$ over the N=0$\rightarrow$1 Coulomb peak. The x-axis has been converted to energy using the lever arm (Suporting Information S). The ground state (GS) and excited states (ES1-4) for the one hole system are labelled. Additional structure is observed for ES2 and ES3, see text. \textbf{c}, Same as (\textbf{b}) for the N=1$\rightarrow$2 Coloumb peak. The inset demonstrates that the ground state and first excited state are resolvable. Each dot represents a single data point. \textbf{d}, Plot of the extracted excited state energies for the one (black) and two (red) hole system. The black dashed line is a straight line fit to the N=1 data for orbital number $\geq$1, highlighting the linear dependence of the excited state energies on the orbital number.}
	\label{fig:ExcitedStates}
\end{figure*} 
In Figure \ref{fig:ShellFilling}g we present the hole orbital spectrum  extracted from the measurements of $E_{add}$ in Figures \ref{fig:ShellFilling}a$-$e. The orbital structure and spin filling for the first six holes is consistent with the expected spin-filling for a parabolically confined two-dimensional quantum dot. The four-fold degeneracy of the 2\textit{$p_{x}$} and 2\textit{$p_{y}$} orbital levels at B=0 demonstrates that the quantum dot has remarkably circular confinement. A key result is the observation of consecutive filling of holes with the same spin orientation ($\downdownarrows$ and $\upuparrows$), which occurs in the 2\textit{p} orbital. Previous studies of silicon hole quantum dots show only alternating spin filling\cite{roddaro2008spin,zwanenburg2009spin,hu2012hole,brauns2016electric}. The results in Figure \ref{fig:ShellFilling}g provide a clear demonstration of the orbital shell spin structure of the first eight holes in a surface-gated silicon quantum dot. In particular, we highlight the observation that holes have spin polarized filling of the 2\textit{p} orbital, analogous to Hund's first rule of orbital shell filling in atomic physics.

We now discuss the spin filling sequence for |B|$\ge2.7$T. The change in slope of $E_{add}(1)$, $E_{add}(2)$ and $E_{add}(3)$ at B=2.7T can be attributed to a magnetic field induced crossing of the 1\textit{s} and 2\textit{p} orbitals, as shown in Figure \ref{fig:ShellFilling}g). We have extracted the orbital effective g-factors $g^{*}$ from Figure \ref{fig:ShellFilling}f to be $g^{*}_{1s}= 1.08$ and $g^{*}_{2p} = 1.39$. By calculating the Zeeman energy at the 1s and 2p crossing we determine the singlet-triplet energy spacing $E_{ST}$ for the two hole dot is 0.2meV.

The change in slope around 2.7T for $E_{add}(4)$ and $E_{add}(5)$ can be attributed to a crossing between the 2\textit{p} orbital and the next highest orbital level. The next highest orbital level above the 2\textit{p} orbital is two-fold degenerate and is occupied by the $7^{th}$ and $8^{th}$ holes (Figure \ref{fig:ShellFilling}g). For  circular 2D confinement the orbital level above 2\textit{p} is expected to be six-fold degenerate. We suspect that the two-fold degenerate orbital above the 2\textit{p} orbital may be result from a loss of circular symmetry of the dot due to for higher hole occupations, or many body effects\cite{ciorga2000addition}.
 
To further study the orbital shell structure, and the nature of the confinement potential we examined the excited state spectrum of the quantum dot. Figure \ref{fig:ExcitedStates}a shows the charge stability diagram when $V_{pulse}$ is increased to 40mV. Increasing $V_{pulse}$ broadens the charge transition window, allowing single hole tunneling to occur via either the ground state or an excited state. The excited state spectrum can be resolved by observing the additional structure of $I_{pulse}$ within broadened charge transition lines.

Figure \ref{fig:ExcitedStates}b shows the excited state spectrum for the dot with single hole occupation. This spectrum is obtained from a high resolution cut of $I_{pulse}$ vs. $V_{G3}$ along the dashed white line labeled (b) in Figure \ref{fig:ExcitedStates}a. The x-axis in Figure \ref{fig:ExcitedStates}b is converted to energy using the lever arm (see Supporting Information) and the ground state is set to zero energy. Peaks in Figure \ref{fig:ExcitedStates}b correspond to the single hole tunneling into different orbital states in the unoccupied quantum dot (0$\rightarrow$1 transition).

The extracted orbital energies are plotted as black circles in Figure \ref{fig:ExcitedStates}d, and show a linear dependence on orbital number. This linearity implies a remarkably parabolic confinement potential. We note that additional structure can be observed in $I_{pulse}$ for the second excited state (ES2) and the third excited state (ES3) in Figure \ref{fig:ExcitedStates}b. This additional structure is likely due to orbital splitting resulting from ellipticity of the dot for higher energy orbitals, consistent with results in Figure \ref{fig:FigureDevice}d.

We now estimate the expected excited state energy scales in order to compare with the experiment. The quantum dot radius was calculated to be $\sim$27nm, by approximating the dot as a parallel plate capacitor and using the charging energy of 12meV for the one to two hole charge occupation. A dot radius of 27nm is smaller than previous silicon MOS hole quantum dots operating in the few hole regime\cite{li2013single,spruijtenburg2013single,yamaoka2017charge}. The expected orbital spacing for a 2D artificial atom with 27nm radius is $\sim$3meV, which is consistent with the measured orbital spectrum in Figure \ref{fig:ExcitedStates}d.

Finally, we investigated the energy spectrum of the two hole quantum dot. We can determine the strength of hole-hole interactions within the quantum dot by comparing the two-hole energy spectrum with the one-hole energy spectrum. Figure \ref{fig:ExcitedStates}c shows the excited state spectrum for the two-hole quantum dot, which is a cut along the dashed white line labeled (c) in Figure \ref{fig:ExcitedStates}a. A key feature of the two hole dot is that the first excited state is now only 0.25meV above the ground state (inset of Figure \ref{fig:ExcitedStates}c), while the separation between excited states remains comparable to the N=1 transition excited state energy separation of $\sim$3meV. The reduction in the spacing between the ground state and first orbital state (3.5meV for the one hole system, and 0.25meV for the two hole system) results from the additional Coulomb interaction energy when one hole already occupies the lowest energy orbital. The observation of a 0.25meV excited state spacing for the two-hole dot is consistent with the 0.2meV Zeeman energy required to induce a singlet-triplet ground state transition in Figure \ref{fig:ShellFilling}e. Based on the change in first orbital energy spacing, we estimate that the hole interaction energy is $\sim$90\% of the orbital energy, which is much larger than in electron systems\cite{elzerman2004excited,yang2012orbital}. This large interaction energy has significant implications for Pauli spin blockade and quantum information applications.

In summary, we have demonstrated a silicon MOS based quantum dot operating in the last hole regime. The orbital level spacing demonstrates that the confinement potential is remarkably parabolic. We have extracted the ground state spin filling for N=1 to 8 holes. The spin shell filing for the first six holes is found to be consistent with predictions for a circular 2D quantum dot. Finally, we determine that strong hole-hole interactions affect the two-hole energy spectrum, resulting in suppression of the singlet-triplet energy spacing. These results highlight the unique physics of 2D hole artificial atoms, and clearly demonstrate that spin properties and energy scales are very different to nanowire and electron artificial atoms\cite{roddaro2008spin,zwanenburg2009spin,hu2012hole,zwanenburg2013silicon}.

\scriptsize
\subsection*{Methods}
\textbf{The Sample:} The device studied in this work was fabricated using a high resistivity natural (001) silicon substrate. The P+ ohmic regions are prepared by boron diffusion. A 5.9nm gate dielectric (SiO$_{2}$) is grown by dry oxidation in the active region of the device. The gate pattern is fabricated using multilayer Al-Al$_{2}$0$_{3}$ gate stack technology\cite{li2013single}. The final stage is a forming gas (95\% N$_{2}$/5\% H$_{2}$) anneal to reduce Si/SiO$_{2}$ interface disorder and enhance low temperature performance. All measurements were performed in a dilution fridge with a base temperature below 30mK.

When operating the device, the reservoir top gate is negatively biased to accumulate a 2D hole system at the Si/SiO$_{2}$ interface below. The quantum dot is defined by positively biasing gates G1, G2, G4, and the C-gate. G3 acts as the dot plunger gate and is operated in the negatively biased regime. It is possible to operate this device in the double dot regime down to the (0,0) charge state, using gate G2 as the second dot's plunger gate. Additionally, in the double do regime we can observe interdot tunneling. See Supporting information for more information on the tunability of this device.

\textbf{Charge Sensor:} The pulse-bias charge sensing method has been extensively described in Refs. [\citenum{elzerman2004excited,yang2012orbital}]. In order to maximize the sensitivity of the charge sensor we use a feedback loop on the sensor gate to keep the sensor at the same conductance as the other gates are swept, as described in Ref. [\citenum{yang2011dynamically}]. We note that the charge transitions signals using $I_{pulse}$ in Figure \ref{fig:FigureDevice}d are sensitive to the dot tunnel rate, hence we have also confirmed the charge occupation by simultaneously measuring the sensor conductance using $I_{sensor}$, which is not sensitive to the tunnel rate. See the Supporting Information for full details.

\textbf{Magneto-spectroscopy:} In order to infer the spin configuariuon of the dot for different hole occupations we measure the spin state of all additional holes relative to the first hole, which we assume is aligned with the in-plane magnetic field $B$. The N=1 spin ground state is assigned as down. The relative spin orientation of the first six holes can be inferred from the data presented in Figures \ref{fig:ShellFilling}a-e, and is well described by the orbital model presented in Figure \ref{fig:ShellFilling}d). Further discussion regarding the spin filling of the 7$^{th}$ and $8^{th}$ holes is provided in the Supporting Information. 

\textbf{Pulse-bias spectroscopy:} The charge stability diagram shown in Figure \ref{fig:ExcitedStates}a is obtained using the same gate bias configuration as the stability diagram in Figure \ref{fig:FigureDevice}b, except the charge transitions are broader due to increased $V_{pulse}$. When sweeping $V_{G3}$ over a broadened charge transition signal (as indicated by the horizontal dashed lines in Figure \ref{fig:ExcitedStates}a), $I_{pulse}$ initially increases as the ground state is pulsed below the reservoir electrochemical potential $\mu_{res}$. $I_{pulse}$ then decays as $V_{G3}$ becomes more negative, since the effective tunnel barrier increases. For sufficiently negative $V_{G3}$ additional excited states become accessible for tunneling, which increases the tunnel rate and causes additional spikes in $I_{pulse}$.


\begin{thebibliography}{34}
	\providecommand{\natexlab}[1]{#1}
	\providecommand{\url}[1]{\texttt{#1}}
	\expandafter\ifx\csname urlstyle\endcsname\relax
	\providecommand{\doi}[1]{doi: #1}\else
	\providecommand{\doi}{doi: \begingroup \urlstyle{rm}\Url}\fi
	
	\bibitem[Loss and DiVincenzo(1998)]{loss1998quantum}
	Daniel Loss and David~P DiVincenzo.
	\newblock Quantum computation with quantum dots.
	\newblock \emph{Physical Review A}, 57\penalty0 (1):\penalty0 120, 1998.
	
	\bibitem[Hanson et~al.(2007)Hanson, Kouwenhoven, Petta, Tarucha, and
	Vandersypen]{hanson2007spins}
	Ronald Hanson, Leo~P Kouwenhoven, Jason~R Petta, Seigo Tarucha, and Lieven~MK
	Vandersypen.
	\newblock Spins in few-electron quantum dots.
	\newblock \emph{Reviews of Modern Physics}, 79\penalty0 (4):\penalty0 1217,
	2007.
	
	\bibitem[Zwanenburg et~al.(2013)Zwanenburg, Dzurak, Morello, Simmons,
	Hollenberg, Klimeck, Rogge, Coppersmith, and Eriksson]{zwanenburg2013silicon}
	Floris~A Zwanenburg, Andrew~S Dzurak, Andrea Morello, Michelle~Y Simmons,
	Lloyd~CL Hollenberg, Gerhard Klimeck, Sven Rogge, Susan~N Coppersmith, and
	Mark~A Eriksson.
	\newblock Silicon quantum electronics.
	\newblock \emph{Reviews of modern physics}, 85\penalty0 (3):\penalty0 961,
	2013.
	
	\bibitem[Veldhorst et~al.(2014)Veldhorst, Hwang, Yang, Leenstra, de~Ronde,
	Dehollain, Muhonen, Hudson, Itoh, Morello, et~al.]{veldhorst2014addressable}
	M~Veldhorst, JCC Hwang, CH~Yang, AW~Leenstra, Bob de~Ronde, JP~Dehollain,
	JT~Muhonen, FE~Hudson, KM~Itoh, A~Morello, et~al.
	\newblock An addressable quantum dot qubit with fault-tolerant
	control-fidelity.
	\newblock \emph{Nature nanotechnology}, 9\penalty0 (12):\penalty0 981--985,
	2014.
	
	\bibitem[Koppens et~al.(2006)Koppens, Buizert, Tielrooij, Vink, Nowack,
	Meunier, Kouwenhoven, and Vandersypen]{koppens2006driven}
	FHL Koppens, Christo Buizert, Klaas-Jan Tielrooij, IT~Vink, KC~Nowack, Tristan
	Meunier, LP~Kouwenhoven, and LMK Vandersypen.
	\newblock Driven coherent oscillations of a single electron spin in a quantum
	dot.
	\newblock \emph{Nature}, 442\penalty0 (7104):\penalty0 766--771, 2006.
	
	\bibitem[Golovach et~al.(2006)Golovach, Borhani, and
	Loss]{golovach2006electric}
	Vitaly~N Golovach, Massoud Borhani, and Daniel Loss.
	\newblock Electric-dipole-induced spin resonance in quantum dots.
	\newblock \emph{Physical Review B}, 74\penalty0 (16):\penalty0 165319, 2006.
	
	\bibitem[Flindt et~al.(2006)Flindt, S{\o}rensen, and Flensberg]{flindt2006spin}
	Christian Flindt, Anders~S S{\o}rensen, and Karsten Flensberg.
	\newblock Spin-orbit mediated control of spin qubits.
	\newblock \emph{Physical review letters}, 97\penalty0 (24):\penalty0 240501,
	2006.
	
	\bibitem[Nowack et~al.(2007)Nowack, Koppens, Nazarov, and
	Vandersypen]{nowack2007coherent}
	KC~Nowack, FHL Koppens, Yu~V Nazarov, and LMK Vandersypen.
	\newblock Coherent control of a single electron spin with electric fields.
	\newblock \emph{Science}, 318\penalty0 (5855):\penalty0 1430--1433, 2007.
	
	\bibitem[Li et~al.(2015)Li, Hudson, Dzurak, and Hamilton]{li2015pauli}
	Ruoyu Li, Fay~E Hudson, Andrew~S Dzurak, and Alexander~R Hamilton.
	\newblock Pauli spin blockade of heavy holes in a silicon double quantum dot.
	\newblock \emph{Nano letters}, 15\penalty0 (11):\penalty0 7314--7318, 2015.
	
	\bibitem[Spruijtenburg et~al.(2013)Spruijtenburg, Ridderbos, Mueller, Leenstra,
	Brauns, Aarnink, van~der Wiel, and Zwanenburg]{spruijtenburg2013single}
	Paul~C Spruijtenburg, Joost Ridderbos, Filipp Mueller, Anne~W Leenstra,
	Matthias Brauns, Antonius~AI Aarnink, Wilfred~G van~der Wiel, and Floris~A
	Zwanenburg.
	\newblock Single-hole tunneling through a two-dimensional hole gas in intrinsic
	silicon.
	\newblock \emph{Applied physics letters}, 102\penalty0 (19):\penalty0 192105,
	2013.
	
	\bibitem[Yamaoka et~al.(2017)Yamaoka, Iwasaki, Oda, and
	Kodera]{yamaoka2017charge}
	Yu~Yamaoka, Kazuma Iwasaki, Shunri Oda, and Tetsuo Kodera.
	\newblock Charge sensing and spin-related transport property of p-channel
	silicon quantum dots.
	\newblock \emph{Japanese Journal of Applied Physics}, 56\penalty0
	(4S):\penalty0 04CK07, 2017.
	
	\bibitem[Betz et~al.(2014)Betz, Gonzalez-Zalba, Podd, and
	Ferguson]{betz2014ambipolar}
	AC~Betz, MF~Gonzalez-Zalba, G~Podd, and AJ~Ferguson.
	\newblock Ambipolar quantum dots in intrinsic silicon.
	\newblock \emph{Applied Physics Letters}, 105\penalty0 (15):\penalty0 153113,
	2014.
	
	\bibitem[Fock(1928)]{fock1928bemerkung}
	Vladimir Fock.
	\newblock Bemerkung zur quantelung des harmonischen oszillators im magnetfeld.
	\newblock \emph{Zeitschrift f{\"u}r Physik A Hadrons and Nuclei}, 47\penalty0
	(5):\penalty0 446--448, 1928.
	
	\bibitem[Darwin(1931)]{darwin1931diamagnetism}
	Charles~Galton Darwin.
	\newblock The diamagnetism of the free electron.
	\newblock In \emph{Mathematical Proceedings of the Cambridge Philosophical
		Society}, volume~27, pages 86--90. Cambridge University Press, 1931.
	
	\bibitem[Dennis et~al.(2002)Dennis, Kitaev, Landahl, and
	Preskill]{dennis2002topological}
	Eric Dennis, Alexei Kitaev, Andrew Landahl, and John Preskill.
	\newblock Topological quantum memory.
	\newblock \emph{Journal of Mathematical Physics}, 43\penalty0 (9):\penalty0
	4452--4505, 2002.
	
	\bibitem[Jones et~al.(2016)Jones, Gyure, Ladd, Fogarty, Morello, and
	Dzurak]{jones2016logical}
	Cody Jones, Mark~F Gyure, Thaddeus~D Ladd, Michael~A Fogarty, Andrea Morello,
	and Andrew~S Dzurak.
	\newblock A logical qubit in a linear array of semiconductor quantum dots.
	\newblock \emph{arXiv preprint arXiv:1608.06335}, 2016.
	
	\bibitem[Szumniak et~al.(2012)Szumniak, Bednarek, Partoens, and
	Peeters]{szumniak2012spin}
	P~Szumniak, S~Bednarek, B~Partoens, and FM~Peeters.
	\newblock Spin-orbit-mediated manipulation of heavy-hole spin qubits in gated
	semiconductor nanodevices.
	\newblock \emph{Physical review letters}, 109\penalty0 (10):\penalty0 107201,
	2012.
	
	\bibitem[Kloeffel et~al.(2013)Kloeffel, Trif, Stano, and
	Loss]{kloeffel2013circuit}
	Christoph Kloeffel, Mircea Trif, Peter Stano, and Daniel Loss.
	\newblock Circuit qed with hole-spin qubits in ge/si nanowire quantum dots.
	\newblock \emph{Physical Review B}, 88\penalty0 (24):\penalty0 241405, 2013.
	
	\bibitem[Pribiag et~al.(2013)Pribiag, Nadj-Perge, Frolov, Van Den~Berg,
	Van~Weperen, Plissard, Bakkers, and Kouwenhoven]{pribiag2013electrical}
	VS~Pribiag, S~Nadj-Perge, SM~Frolov, JWG Van Den~Berg, I~Van~Weperen,
	SR~Plissard, EPAM Bakkers, and LP~Kouwenhoven.
	\newblock Electrical control over single hole spins in nanowire quantum dots.
	\newblock \emph{arXiv preprint arXiv:1302.2648}, 2013.
	
	\bibitem[Maurand et~al.(2016)Maurand, Jehl, Kotekar-Patil, Corna, Bohuslavskyi,
	Lavi{\'e}ville, Hutin, Barraud, Vinet, Sanquer, and
	Franceschi]{maurand2016cmos}
	R~Maurand, X~Jehl, D~Kotekar-Patil, A~Corna, H~Bohuslavskyi, R~Lavi{\'e}ville,
	L~Hutin, S~Barraud, M~Vinet, M~Sanquer, and Silvano~De Franceschi.
	\newblock A cmos silicon spin qubit.
	\newblock \emph{Nature communications}, 7:\penalty0 13575, 2016.
	
	\bibitem[Bulaev and Loss(2005)]{bulaev2005spin}
	Denis~V Bulaev and Daniel Loss.
	\newblock Spin relaxation and anticrossing in quantum dots: Rashba versus
	dresselhaus spin-orbit coupling.
	\newblock \emph{Physical Review B}, 71\penalty0 (20):\penalty0 205324, 2005.
	
	\bibitem[Keane et~al.(2011)Keane, Godfrey, Chen, Fricke, Klochan, Burke,
	Micolich, Beere, Ritchie, Trunov, et~al.]{keane2011resistively}
	ZK~Keane, MC~Godfrey, JCH Chen, S~Fricke, O~Klochan, AM~Burke, AP~Micolich,
	HE~Beere, DA~Ritchie, KV~Trunov, et~al.
	\newblock Resistively detected nuclear magnetic resonance in n-and p-type gaas
	quantum point contacts.
	\newblock \emph{Nano letters}, 11\penalty0 (8):\penalty0 3147--3150, 2011.
	
	\bibitem[Koppens et~al.(2008)Koppens, Nowack, and Vandersypen]{koppens2008spin}
	FHL Koppens, KC~Nowack, and LMK Vandersypen.
	\newblock Spin echo of a single electron spin in a quantum dot.
	\newblock \emph{Physical review letters}, 100\penalty0 (23):\penalty0 236802,
	2008.
	
	\bibitem[Viola and Lloyd(1998)]{viola1998dynamical}
	Lorenza Viola and Seth Lloyd.
	\newblock Dynamical suppression of decoherence in two-state quantum systems.
	\newblock \emph{Physical Review A}, 58\penalty0 (4):\penalty0 2733, 1998.
	
	\bibitem[Hu et~al.(2012)Hu, Kuemmeth, Lieber, and Marcus]{hu2012hole}
	Yongjie Hu, Ferdinand Kuemmeth, Charles~M Lieber, and Charles~M Marcus.
	\newblock Hole spin relaxation in ge-si core-shell nanowire qubits.
	\newblock \emph{Nature nanotechnology}, 7\penalty0 (1):\penalty0 47--50, 2012.
	
	\bibitem[Higginbotham et~al.(2014)Higginbotham, Larsen, Yao, Yan, Lieber,
	Marcus, and Kuemmeth]{higginbotham2014hole}
	Andrew~P Higginbotham, Thorvald~Wadum Larsen, Jun Yao, Hao Yan, Charles~M
	Lieber, Charles~M Marcus, and Ferdinand Kuemmeth.
	\newblock Hole spin coherence in a ge/si heterostructure nanowire.
	\newblock \emph{Nano letters}, 14\penalty0 (6):\penalty0 3582--3586, 2014.
	
	\bibitem[Yang et~al.(2012)Yang, Lim, Lai, Rossi, Morello, and
	Dzurak]{yang2012orbital}
	CH~Yang, WH~Lim, NS~Lai, A~Rossi, A~Morello, and AS~Dzurak.
	\newblock Orbital and valley state spectra of a few-electron silicon quantum
	dot.
	\newblock \emph{Physical Review B}, 86\penalty0 (11):\penalty0 115319, 2012.
	
	\bibitem[Elzerman et~al.(2004)Elzerman, Hanson, Willems~van Beveren,
	Vandersypen, and Kouwenhoven]{elzerman2004excited}
	JM~Elzerman, R~Hanson, LH~Willems~van Beveren, LMK Vandersypen, and
	LP~Kouwenhoven.
	\newblock Excited-state spectroscopy on a nearly closed quantum dot via charge
	detection.
	\newblock \emph{Applied physics letters}, 84\penalty0 (23):\penalty0
	4617--4619, 2004.
	
	\bibitem[Ciorga et~al.(2000)Ciorga, Sachrajda, Hawrylak, Gould, Zawadzki,
	Jullian, Feng, and Wasilewski]{ciorga2000addition}
	M~Ciorga, AS~Sachrajda, Pawel Hawrylak, C~Gould, Piotr Zawadzki, S~Jullian,
	Y~Feng, and Zbigniew Wasilewski.
	\newblock Addition spectrum of a lateral dot from coulomb and spin-blockade
	spectroscopy.
	\newblock \emph{Physical Review B}, 61\penalty0 (24):\penalty0 R16315, 2000.
	
	\bibitem[Roddaro et~al.(2008)Roddaro, Fuhrer, Brusheim, Fasth, Xu, Samuelson,
	Xiang, and Lieber]{roddaro2008spin}
	Stefano Roddaro, Andreas Fuhrer, Patrik Brusheim, Carina Fasth, HQ~Xu, Lars
	Samuelson, J~Xiang, and CM~Lieber.
	\newblock Spin states of holes in ge/si nanowire quantum dots.
	\newblock \emph{Physical review letters}, 101\penalty0 (18):\penalty0 186802,
	2008.
	
	\bibitem[Zwanenburg et~al.(2009)Zwanenburg, van Rijmenam, Fang, Lieber, and
	Kouwenhoven]{zwanenburg2009spin}
	Floris~A Zwanenburg, Cathalijn~EWM van Rijmenam, Ying Fang, Charles~M Lieber,
	and Leo~P Kouwenhoven.
	\newblock Spin states of the first four holes in a silicon nanowire quantum
	dot.
	\newblock \emph{Nano letters}, 9\penalty0 (3):\penalty0 1071--1079, 2009.
	
	\bibitem[Brauns et~al.(2016)Brauns, Ridderbos, Li, Bakkers, and
	Zwanenburg]{brauns2016electric}
	Matthias Brauns, Joost Ridderbos, Ang Li, Erik~PAM Bakkers, and Floris~A
	Zwanenburg.
	\newblock Electric-field dependent g-factor anisotropy in ge-si core-shell
	nanowire quantum dots.
	\newblock \emph{Physical Review B}, 93\penalty0 (12):\penalty0 121408, 2016.
	
	\bibitem[Li et~al.(2013)Li, Hudson, Dzurak, and Hamilton]{li2013single}
	Ruoyu Li, Fay~E Hudson, Andrew~S Dzurak, and Alexander~R Hamilton.
	\newblock Single hole transport in a silicon metal-oxide-semiconductor quantum
	dot.
	\newblock \emph{Applied Physics Letters}, 103\penalty0 (16):\penalty0 163508,
	2013.
	
	\bibitem[Yang et~al.(2011)Yang, Lim, Zwanenburg, and
	Dzurak]{yang2011dynamically}
	CH~Yang, WH~Lim, FA~Zwanenburg, and AS~Dzurak.
	\newblock Dynamically controlled charge sensing of a few-electron silicon
	quantum dot.
	\newblock \emph{AIP Advances}, 1\penalty0 (4):\penalty0 042111, 2011.
	
\end{thebibliography}



\subsection*{Acknowledgments}
This work was funded by the Australian Research Council (CE11E0001017  and DP150100237) and the US Army Research Office (W911NF-13-1-0024). Devices were made at the NSW node of the Australian National Fabrication Facility. We thank D. Miserev, O. Sushkov, and D. Q. Wang for helpful discussions.

\subsection*{Author Contributions}
S.D.L and R.L performed the experiments and F.E.H and M.V fabricated the device. R.L, S.D.L, C.H.Y, and A.R.H designed the experiments. S.D.L and R.L analyzed the results and S.D.L, R.L, C.H.Y, A.S.D and A.R.H contributed to discussions. S.D.L wrote the manuscript with help from all co-authors.

\subsection*{Additional Information}
The authors have provided Supporting Information that contains the following details; A confirmation of absolute charge occupation of the dot, independent of the tunnel rate; Tuning the dot to reservoir tunnel rate and operation of the device as a double quantum dot; Lever arm and hole temperature calculation; and Spin filling of the $7^{th}$ and $8^{th}$ holes.

\subsection*{Competing financial interests}
The authors declare no competing financial interests.

\end{multicols}

\end{document}